\documentstyle[12pt]{article}

\title{\bf Einsteinian blunders}

\author{Domingos Soares \\  ~\\ Physics Department\\
Federal University of Minas Gerais \\   
Belo Horizonte, Brazil}

\date{March 7, 2005}

\begin{document}
\maketitle

\hfill {\it `We are certainly not to relinquish the evidence of experiments} 

\hfill {\it for the sake of dreams and vain fictions of our own devising.' }\\

\hfill Mathematical Principles of Natural Philosophy,

\hfill  Book III --- I. Newton, 1687\\

\begin{abstract}
The World Year of Physics 2005 celebrates Einstein 1905. Too much 
celebration of a single character may be hazardous as an example for the 
younger generation. With such a motto in mind, I comment on some episodes 
in Einstein's scientific career that are at some extent characterized by 
what could be dubbed {\it blundering behavior}. The final purpose is 
obviously to {\it humanize} the personage as opposed to the current trend 
of {\it deification}.
\end{abstract}

\bigskip\bigskip\bigskip

\section{Introduction}
{\it ``The World Year of Physics 2005 is an United Nations endorsed 
international celebration of physics. Events throughout the year will 
highlight the vitality of physics and its importance in the coming millennium, 
and will commemorate the pioneering contribution of Albert Einstein in 1905."
}

This is the opening statement that appears 
in the electronic page \\{\tt http://www.physics2005.org}, 
which is dedicated to the World Year of 
Physics. The idea seems to be, first, to celebrate physics, and, second, 
to commemorate Einstein. However, one sees without much effort that 
already, in the beginning of the year, there has been too much talking and 
writing on Einstein, with a noticeable bias to scientific {\it idolatry}, 
an unimaginable feature in science. We, scientists, are supposed to respect 
Nature as the sole source of inspiration for our activities both in the 
experimental and theoretical realms.

It is beyond of doubt that 1905 was Einstein's {\it annus mirabilis}. In 
that year the world witnessed the publication of three masterpieces in the 
literature of contemporary physics.They were the work on Brownian motion, 
establishing the reality of atoms, the work on the photoelectric effect 
establishing the quanta of radiation, and the special theory of relativity. 
There was though an antecedent of such a high moment in science: the year 
1666 is often remembered as Isaac Newton's {\it annus mirabilis}. From 
1665 to 1667 he also opened the doors to three new areas of scientific 
research, namely, he laid down the foundation of differential and 
integral calculus, he developed the theory of colors, and put forward 
his theory of gravitation. The publication in 1687 of his 
{\it Mathematical Principles of Natural Philosophy} marked the beginning 
of a new era in the scientific endeavor. There is a 
clear parallel with Einstein's contribution to modern science.

In the following three sections I comment on aspects of Einstein's scientific 
life that are often seen with {\it respectful acceptance}, in spite of 
being bad examples of scientific manners. 

\section{The cosmological constant }
In an important paper published in the Annals of the Royal Prussian 
Acad\-e\-my of Sciences in 1917 entitled {\it ``Kosmologische 
Betrachtungen zur Allgemeinen Relativit\"atstheorie''}, i.e., 
{\it "Cosmological Considerations on the General Theory of Relativity''}, 
Einstein inaugurated the era of 
modern cosmology applying his ideas from General Relativity to the 
universe as a whole. Towards that aim, he abandoned his original 
field equations in favor of a new law in which there was an additional 
constant term that represented a repulsive gravitational potential (Rindler 
2006, p. 304). The term gives a small repulsion near the origin but increased directly 
proportional to the distance until counterbalance the gravitational attraction 
between masses. His intention was obviously 
obtain a {\it static} model. Remember that at that time even galaxies 
were not known as independent cosmological entities. Only in the late 
1920s, with the work of the astronomer Edwin P. Hubble the existence of 
galaxies came to be definitely proved. The solutions he had initially obtained 
with the application of the original field equations were unstable for 
gravitational collapse. The modification preserved the general covariance 
of the theory and solved the instability problem (North 1990, chapter 5). 
The constant became known as the {\it cosmological constant}, and is until 
today the matter of much debate.

I describe now some facts that led to the first {\it Einsteinian blunder.} 
It turns out to be a {\it double-blunder}, as I suggest below.

The introduction of the cosmological constant led to a great debate on many 
aspects of the new horizons opened up by General Relativity concerning the 
universe. A infinite model has obvious boundary condition problems, something 
that was recognized even in the context of a Newtonian cosmology
(see Harrison 2000, chapter 16). With the cosmological constant, Einstein 
solved all problems by introducing a {\it finite, spatially closed} and 
{\it static} model, the latter feature being a result of Einstein's --- and 
of most scientists at the time --- belief concerning the physical world.

Nevertheless, Einstein came later to reject his own modification of the field 
equations. North (1990, p. 86) quotes that already by 1919 Einstein considered 
that the introduction of the constant was {\it ``gravely detrimental to the 
formal beauty of the theory''}; he considered it as an {\it ad hoc} 
addition to the field equations. Later on, he was further led to such a 
rejection by two new developments: on the observational side, Hubble's work 
on the redshift-distance relation for galaxies was being interpreted as an 
indication of an expanding universe\footnote{It is worthwhile to note at 
this point that the interpretation of Hubble's redshift-distance relation 
as indicative of an expanding universe is only true when one takes for 
granted that the underlying theory under consideration, i.e., 
General Relativity in modern cosmology, is true. This is still a matter of 
debate since present cosmological models have led to a variety of 
hypotheses concerning the matter-energy content of the universe,
such as {\it baryonic dark matter, non-baryonic dark matter} and 
the yet more mysterious {\it dark energy}. None of these have been so far 
proved to exist by any experimental or observational means.}
 --- no need of static solutions ---, 
and on the theoretical side, the 1922 solution of the field 
equations by the Russian Aleksandr Friedmann and the 1927 solution by the 
Belgian Georges Lema\^\i tre which allowed for expanding models. 

George Gamov (1970) tells the now legendary story that Einstein once has said 
to him that the cosmological constant was {\it ``my biggest 
blunder''} \footnote{This of course entirely justifies the title of the 
present article: if Einstein admits his ``biggest'' {\it blunder}, that 
implies the existence of the ``smallest'', and a whole gradation of {\it 
blunders} in between.}.

But why a double-blunder? Einstein rejected the cosmological constant based 
on what he found to be physical and aesthetic inconsistencies that resulted 
from its adoption. Here he saw a blunder, his biggest one. On the other hand, 
from the strict theoretical and formal point of views, General Relativity 
is in fact {\it enriched} by the addition of the new term, while still 
keeping its features as a viable general covariant theory of gravitation. 
And that is where the {\it double} character comes from. The simple fact 
of abandoning it constitutes a blunder after a blunder. This is 
also suggested by Norh's arguments (North 1990, p. 86), who writes that 
{\it ``he finally discarded the term in 1931, and in doing so deliberately 
restricted the generality of his theory.''}

Recent claims, from the late 1990s and on, of a {\it accelerating} expanding 
universe have led to the resurrection of the cosmological constant, which  
would give the {\it cosmic repulsion} responsible for the acceleration. 
This idea and other variants became a strong feature of modern cosmology. 
The present {\it status quo} of modern cosmology is not though free of 
opposition. An example of that has recently materialized in 
{\it An Open Letter to the Scientific Community} (Lerner 2004).      

\section{The 1919 solar eclipse }
In a short biography (Bernstein 1976), Einstein's reactions to the 
scientific results obtained from the solar eclipse of 1919 are described. 
The main issue was light bending by a gravitational source, and the occasion 
was most appropriate for the observational tests.

Two astronomical expeditions, one in Brazil 
and another in the African coast, were organized by Sir Arthur Eddington, 
a renowned scientist at the time, in order to 
measure the stellar positions around the solar disk during the total eclipse 
of May 29, 1919. Ilse Rosenthal-Schneider, Einstein's student, tells that 
Einstein's first reaction to the news that the measurements pointed to 
an agreement with General Relativity predictions for the light bending was: 
{\it ``--- I knew it was correct''}. 

She asked him: {\it ``--- What would it be if your prediction was not 
confirmed?''.} He replied: {\it ``--- Da k\"onnt' mir halt der 
liebe Gott leid tun, die Theorie stimmt doch.''} Or, {\it ``--- Then 
I would be sorry for the good Lord, but the Theory is correct.''}

Is this an acceptable reaction of a theorist when confronted with experiments 
or observations that are relevant to his theory? Certainly not. 

\section{Einstein meets Hubble }
The protagonist here is another Einstein --- Elsa --- Einstein's second 
wife. She is sometimes featured as a woman of somewhat {\it faint} character 
(see Pais 1983). The story appears in many sources. The one I quote here 
is from the probably best biography of the great extragalactic astronomer 
Edwin Powell Hubble (Christianson 1995), the man that successfully proved the 
existence of external galaxies and would be awarded the Nobel prize in Physics 
in the early 1950s. It did not happen due to his premature death in 1953 
(for a short account, see Soares 2001).

Einstein's visit to the institutes of Caltech, in early 1931, was motivated 
by his curiosity on the work in mathematical physics done by Richard Tolman, 
who was working on relativity, and on the observational work by Hubble 
at the Mount Wilson Observatory. 

He and wife made their first trip to the mountain, where the Observatory was 
located, in mid-February. They were accompanied by Hubble and others.

They visit all the installations in the Observatory, including the 100-inch 
dome, which houses the Hooker telescope --- then the largest telescope of the 
world ---, where most of Hubble's work on extragalactic astronomy was being 
conducted. 

Hubble's biographer writes (p. 206): {\it ``When Elsa Einstein, 
who seemed always to be in the defensive, was told that the giant Hooker 
telescope was essential for determining the universe's structure, she is 
said to have replied, `Well, well, my husband does that on the back of an old 
envelope.' "} As in the previous section, one sees here the diminution of 
the relevance of experimental (strictly speaking, observational) science. 

One could argue that this is not a legitimate Einsteinian blunder because it 
was Mrs. Elsa's mouth that has spoken out the words. There are two 
counter-arguments against such a claim. The {\it weak} and the {\it strong}  
arguments. The weak one is just a play on words and goes like this: "Elsa is 
Einstein therefore it is an Einsteinian blunder". The strong argument is that 
the episode appears very frequently in Einstein's biographies and in 
writings of various nature about both Einstein and Hubble. It is an 
Einsteinian feature. As such, it might with justice be included in the 
gallery of authentic Einsteinian blunders. 

\section{Concluding remarks}
It is understandable that amongst us, physicists and astronomers, there is 
frequently almost an adoration of Albert Einstein. He is without doubt the 
greatest scientist of the Twentieth century. Such an involuntary worship is 
everywhere: the most celebrated Einstein's biography, namely, that by 
Abraham Pais (Pais 1983) is also contaminated. He adopts the usual trend of 
skipping uncomfortable details of Einstein's personal and scientific life  
(Soares 2003). 

The reader certainly noticed that none of the above-mentioned stories refers 
to Einstein's {\it annus mirabilis} works but are at some extent related 
to General Theory of Relativity, which was developed ten years later. 
The explanation is simple. It is the result of a {\it selection effect}, 
given that the author of the present article is an extragalactic astronomer. 
That is to say, it does not means that there are not Einsteinian blunders 
related to that period. They can be mined, for example, in Abraham Pais' 
book. Not without some effort, it should be added, as implied by the first  
paragraph, 

And what about the atomic bomb? Certainly it cannot be classified as an 
Einsteinian blunder, in spite of Einstein's deep involvement with the 
issue (especially on the political side, see 
Pais 1983 for details). The atomic bomb is rather the {\it world biggest 
blunder. }

Scientific impartiality excludes, by definition, worshiping and the cult 
of personality. To err is {\it human} and so has Einstein erred in 
many occasions. This is the plain message to the younger generation of 
students and scientists.

Finally, young and old, let us all remember the famous Brazilian playwright 
Nelson Rodrigues that always used to say that {\it ``any unanimity is 
stupid.''} Definitely right.

\section{References}
\begin{description}
\item Bernstein, J. 1976, {\it Einstein}, Viking Press, New York
\item Christianson, G.E. 1995, {\it Edwin Hubble: Mariner of the Nebulae}, The 
University of Chicago Press, Chicago
\item Gamov, G. 1970, {\it My World Line}, Vicking Press, New York
\item Harrison, E.R. 2000, {\it Cosmology, the Science of the Universe}, 
Cambridge University Press, Cambridge
\item Lerner, E. 2004, {\it Bucking the Big Bang}, New Scientist, 182 (2448), 
20 (also at {\tt http://www.cosmologystatement.org})
\item North, J.D. 1990, {\it The Measure of the Universe: A History of Modern 
Cosmology}, Dover Publications, Inc., New York
\item Pais, A. 1983, {\it Subtle Is the Lord: The Science and the Life of 
Albert Einstein}, Oxford University Press, Oxford
\item Rindler, W. 2006, {\it Relativity: Special, General, and Cosmological}, Oxford University Press, New York
\item Soares, D.S.L. 2001, {\it Hubble's Nobel Prize}, The Journal of the 
Royal Astronomical Society of Canada, 95, 10 (also at\\
{\tt http://www.fisica.ufmg.br/\char126 dsoares/hubble/hubble.html})
\item Soares, D.S.L. 2003, {\it Subtle is Abraham Pais...}, \\
{\tt http://www.fisica.ufmg.br/\char126 dsoares/abpais/abpais.htm}
\end{description}

\end{document}